\documentclass[12pt, preprint,aps,prb,superscriptaddress]{revtex4-2}
\usepackage{hyperref}
\hypersetup{
    colorlinks=true,
    linkcolor=blue,
    urlcolor=blue,
    citecolor=blue
}
\usepackage{standalone}
\usepackage{orcidlink}
\usepackage[autostyle]{csquotes}
\usepackage[utf8x]{inputenc}
\usepackage[T1]{fontenc}
\usepackage{natbib}
\usepackage{bpchem}
\usepackage{siunitx}
\usepackage{graphicx}
\usepackage{amsmath,amssymb,amstext}
\usepackage{breqn}
\usepackage{multirow}
\usepackage{float}
\usepackage{endnotes}
\usepackage{tabularx}
\usepackage{subcaption}
\usepackage{url}
\usepackage{setspace}

\makeatletter
\preto\maketitle{%
  \begingroup\lccode`~=`,
  \lowercase{\endgroup
  \let\saved@breqn@active@comma~
  \let~}\active@comma 
}
\appto\maketitle{%
  \begingroup\lccode`~=`,
  \lowercase{\endgroup
  \let~}\saved@breqn@active@comma % 
}
\makeatother

\begin{document}

\title{Accelerated ultrafast demagnetization of an interlayer-exchange-coupled Co/Mn/Co trilayer}

\author{Jendrik G\"ordes\orcidlink{0000-0003-4321-8133}}
\affiliation{Institut f\"ur Experimentalphysik, Freie Universit\"at Berlin, Arnimallee 14, 14195 Berlin, Germany}
\author{Ivar Kumberg\orcidlink{0000-0002-3914-0604}}
\affiliation{Institut f\"ur Experimentalphysik, Freie Universit\"at Berlin, Arnimallee 14, 14195 Berlin, Germany}
\author{Chowdhury S. Awsaf\orcidlink{0009-0007-4709-6168}}
\affiliation{Institut f\"ur Experimentalphysik, Freie Universit\"at Berlin, Arnimallee 14, 14195 Berlin, Germany}
\author{Marcel Walter\orcidlink{0009-0003-4145-0358}}
\affiliation{Institut f\"ur Experimentalphysik, Freie Universit\"at Berlin, Arnimallee 14, 14195 Berlin, Germany}
\author{Tauqir Shinwari\orcidlink{0000-0002-3876-3201}}
\affiliation{Institut f\"ur Experimentalphysik, Freie Universit\"at Berlin, Arnimallee 14, 14195 Berlin, Germany}
\author{Sangeeta Thakur\orcidlink{0000-0003-4879-5650}}
\affiliation{Institut f\"ur Experimentalphysik, Freie Universit\"at Berlin, Arnimallee 14, 14195 Berlin, Germany}
\author{Sangeeta Sharma\orcidlink{0000-0001-6914-1272}}
\affiliation{Institut f\"ur Experimentalphysik, Freie Universit\"at Berlin, Arnimallee 14, 14195 Berlin, Germany}
\affiliation{Max-Born-Institut f\"ur Nichtlineare Optik und Kurzzeitspektroskopie, Max-Born Stra{\ss}e 2a, 12489 Berlin}
\author{Christian Sch\"u{\ss}ler-Langeheine\orcidlink{0000-0002-4553-9726}}
\affiliation{Helmholtz-Zentrum Berlin f\"ur Materialien und Energie, Albert-Einstein-Stra{\ss}e 15, 12489 Berlin}
\author{Niko Pontius\orcidlink{0000-0002-5658-1751}}
\affiliation{Helmholtz-Zentrum Berlin f\"ur Materialien und Energie, Albert-Einstein-Stra{\ss}e 15, 12489 Berlin}
\author{Wolfgang Kuch\orcidlink{0000-0002-5764-4574}}
\email{Correspondence and requests for materials should be addressed to W.K. (email: kuch@physik.fu-berlin.de)}
\affiliation{Institut f\"ur Experimentalphysik, Freie Universit\"at Berlin, Arnimallee 14, 14195 Berlin, Germany}

\date{\today} 

\renewcommand{\thefigure}{\arabic{figure}}
\renewcommand{\theequation}{\arabic{equation}}
\renewcommand{\thetable}{ \Roman{table}}
%\doublespacing
\maketitle

\section*{Abstract}
We investigate the ultrafast magnetization dynamics of an interlayer-exchange-coupled Co/Mn/Co trilayer system after excitation with an ultrafast optical pump. We probe element- and time-resolved ferromagnetic order by X-ray magnetic circular dichroism in resonant reflectivity. We observe an accelerated Co demagnetization time in the case of weak total parallel interlayer coupling at 9.5 ML Mn thickness for antiparallel alignment of both Co layers compared to parallel alignment as well as for parallel alignment in the case of strong parallel interlayer coupling at 11 ML of Mn. From \textit{ab initio} time-dependent density functional theory calculations, we conclude that optically induced intersite spin transfer of spin-polarized electrons from Co into Mn acts as a decay channel to enhance and accelerate ultrafast demagnetization. This spin transfer can only take place in case of a collinear Mn spin structure. We argue that this is the case for antiparallel alignment of both Co layers at 9.5 ML Mn thickness and parallel alignment in case of 11 ML of Mn. Our results point out that an antiferromagnetic spacer layer and its spin structure have a significant effect on the magnetization dynamics of adjacent ferromagnetic layers. Our findings provide further insight into fundamental mechanisms of ultrafast demagnetization and may lead to improve dynamics in multilayered systems for faster optical switching of magnetic order.

\section{Introduction}
\label{Introduction}
Driving spin dynamics of magnetic thin films by ultrashort laser pulses provides the opportunity to manipulate the magnetic order on the femtosecond scale and has gathered interest since pioneering experiments by Beaurepaire \textit{et al}. on Ni in 1996 \cite{Beaurepaire.1996,Hohlfeld.1997, Scholl.1997,Koopmans.2005,Kirilyuk.2010,Walowski.2016}. A number of different theoretical approaches has been employed to explain ultrafast magnetization dynamics, e.g. via superdiffusive spin currents \cite{Malinowski.2008,Battiato.2010,Melnikov.2011,Rudolf.2012,Eschenlohr.2013,Kumberg.2020}, magnons \cite{Carpene.2008, Iacocca.2019} or (Elliott-Yafet, electron-electron, electron-phonon) spin-flip scattering \cite{Stamm.2007,Koopmans.2010,Sultan.2012}. 

Recently, a laser-induced spin-selective excitation of electrons between magnetic sublattices, so-called optically induced intersite spin transfer (OISTR) \cite{Dewhurst.2018}, has emerged as a mechanism to enable extremely fast all-optical manipulation of magnetic order. Due to its purely optical nature, spin dynamics can be induced on timescales shorter than the exchange interaction. Since its theoretical prediction, numerous experiments have confirmed the presence of OISTR, starting by using time-resolved magnetic circular dichroism to observe a faster demagnetization of ferromagnetic (FM) Ni in Ni/Pt multilayers compared to a pure Ni layer \cite{Siegrist.2019}. Later, studies revealed the effect of intersite spin transfer on the enhanced demagnetization of a CoPt alloy in comparison to pure Co \cite{Willems.2020}, as well as a transient FM state of antiferromagnetic (AFM) Mn in Co/Mn multilayers \cite{Golias.2021}. Using transverse magneto-optical Kerr effect in the extreme ultraviolet region, the OISTR effect could be verified in an FeNi alloy and Heusler half-metal \cite{Ryan.2023, Moller.2024}. Very recently, time- and angle-resolved photoemission spectroscopy revealed OISTR to induce a spin-selective charge flow between surface and bulk states in metallic Gd during laser-driven demagnetization, causing a transient increase of the bulk-band exchange splitting \cite{Bobowski.2024}. 

The interface plays an important role for the magnetization dynamics of a FM layer as the FM can be influenced either by direct or indirect interlayer exchange coupling. Kumberg \textit{et al}. \cite{Kumberg.2020} observed an accelerated demagnetization of a FM Ni or Co layer in the presence of antiferromagnetic order in an adjacent NiMn layer, which was attributed to superdiffusive spin currents between the AFM and the FM layer. In the case of a Ni/Ru/Fe FM/spacer/FM trilayer system \cite{Rudolf.2012}, where the magnetization of both FM layers can either be parallel or antiparallel to each other, a transiently enhanced magnetization (parallel) or reduced magnetization (antiparallel) of Fe was observed after an optical pump pulse and attributed to superdiffusive spin currents. Interestingly, a faster demagnetization of Ni was measured in the case of antiparallel alignment but not discussed further. A faster and larger amount of demagnetization for antiparallel alignment of two out-of-plane magnetized FM layers was also observed in the $\mathrm{[CoPt]_n}$/Ru/$\mathrm{[CoPt]_n}$ \cite{Malinowski.2008} multilayer system. It was argued that a direct transfer of spin angular momentum between both CoPt layers takes place via superdiffusive spin currents, which can travel through the metallic Ru spacer but are blocked by insulating AFM NiO.

In this article, we report on the influence of an AFM Mn spacer layer on the magnetization dynamics of adjacent FM Co layers after excitation by an ultrashort laser pulse. For the epitaxial Co/Mn/Co thin-film system, a combination of direct exchange coupling through the spin structure of the AFM layer, orange-peel coupling and Ruderman--Kittel--Kasuya--Yosida (RKKY)-type coupling results in a Mn-thickness-dependent oscillatory interlayer coupling energy between the two Co layers across the Mn layer \cite{Zhang.2014}. By growing the Mn layer as a wedge, different coupling regimes are accessible. We use resonant soft-X-ray magnetic circular dichroism in reflectivity (R-XMCD) to probe the magnetization dynamics after excitation with an ultrashort infrared pulse with elemental and time resolution. Remarkably, we observe an accelerated demagnetization not only in the case of initial antiparallel alignment of both Co layers, in agreement with results from \cite{Malinowski.2008} and \cite{Rudolf.2012}, but also for parallel alignment in the case of a Mn thickness leading to strong parallel interlayer coupling. Applying \textit{ab initio} time-dependent density functional theory (TD-DFT) calculations we identify this behavior to originate from the OISTR effect as the dominant mechanism due to an additional decay channel from a spin-selective transfer of spin-polarized electrons from Co into Mn. Depending on the alignment of both FM layers with respect to the sign of the interlayer coupling by direct exchange through the Mn layer, the Mn spin structure is either collinear or twisted, resulting in OISTR or no or less OISTR, respectively. 

\section{Experiment}
\label{Experiment}
The sample consists of Cu(001)/8 ML Co/7--11 ML Mn(wedge)/20 ML Co/6 ML Ni and was grown under ultra-high vacuum conditions via molecular beam epitaxy. Before evaporation, the Cu(001) single crystal was cleaned by Ar$^+$ sputtering and annealed at a temperature of 850\,K for 30\,min. Substrate cleanliness was checked by Auger electron spectroscopy (AES) and low-energy electron diffraction (LEED). The base pressure of the UHV chamber was around $5 \times 10^{-10}$\,mbar, while the pressure during layer deposition was kept in the lower $10^{-9}$\,mbar region. During evaporation, a quartz crystal microbalance (QCM) and oscillations from medium-energy electron diffraction (MEED) were used to monitor and ensure layer-by-layer growth of the Co bottom layer and the first 7 ML of Mn. For the last 4 ML of Mn, a shutter was placed in front of the moving substrate to grow Mn in a wedge. Deposition of Co and Ni top layers was monitored by QCM since there are no MEED oscillations observable for deposition on the Mn wedge. The layer thickness was additionally checked via AES after deposition of each layer. Magnetic properties of the sample were measured with longitudinal magneto-optical Kerr effect (L-MOKE) at different positions along the wedge. Previous experiments \cite{Zhang.2014} had shown a Mn-thickness-dependent oscillatory interlayer coupling energy between the two ferromagnetic Co layers across the antiferromagnetic Mn layer. Depending on the Mn thickness, either a ferromagnetic coupling or an antiferromagnetic coupling between both Co layers was observed. The reason for this oscillatory behavior was concluded to be from the combination of direct exchange interaction as well as RKKY interaction \cite{1745}, with different periods of oscillation: While the oscillation period of the direct exchange interaction is 2 ML, about 5--6 ML period is expected from the RKKY interaction \cite{Zhang.2014}.  In addition, the constant, non-oscillatory magnetostatic interlayer coupling due to correlated interface roughnesses \cite{Neel62,Dieny91} adds a parallel coupling energy.

\begin{figure}[hbt]
	\includegraphics[width=0.55\columnwidth]{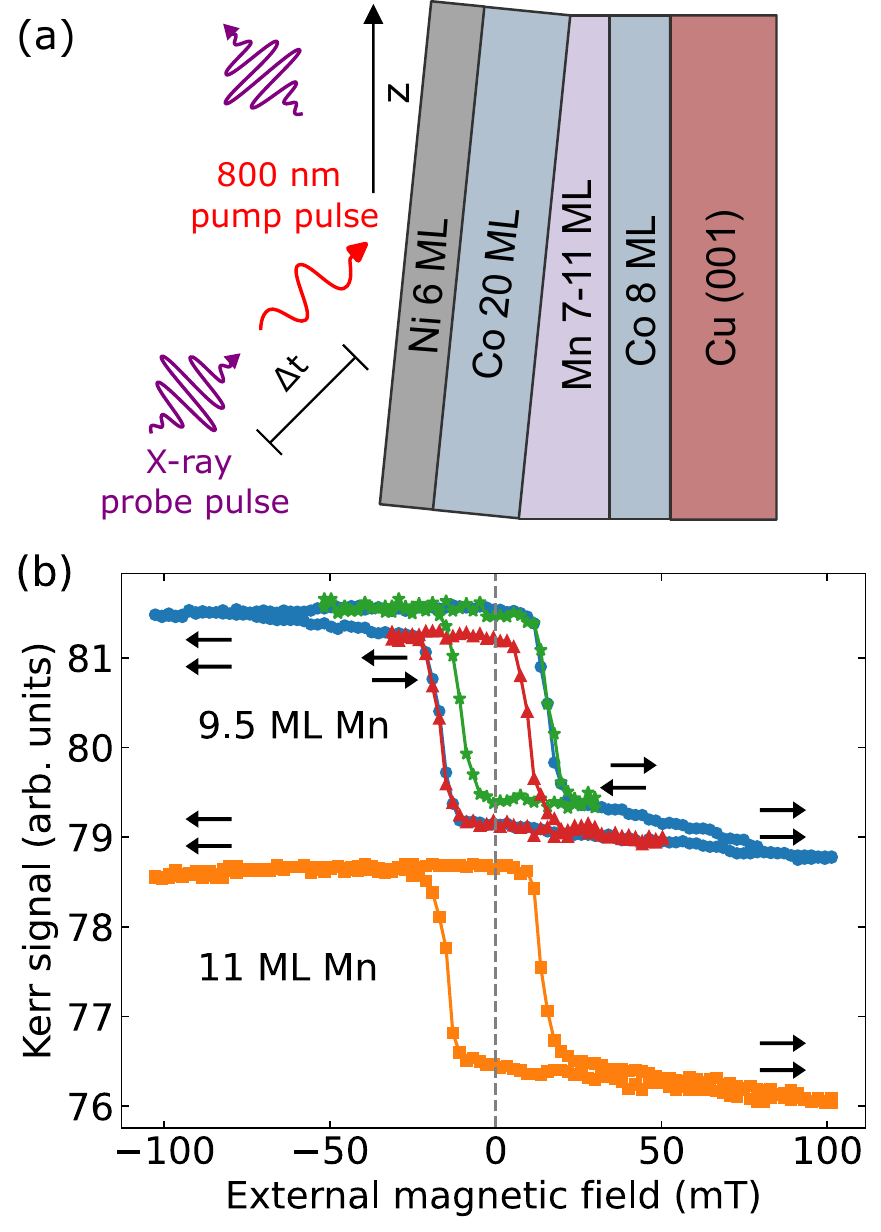}
\caption{(a): Sketch of the experiment. Sample geometry of the Cu(001)/Co/Mn/Co/Ni layer system. (b) The hysteresis loop in the SC regime (orange) shows a one-step magnetization reversal (indicated by black arrows), while in the weak coupling regime (blue) first the top Co layer switches and then the bottom Co layer. Minor loops taken after positive (red) and negative saturation (green) in the WC regime exhibit a positive horizontal shift indicating parallel interlayer coupling.}	
	\label{Fig1}
\end{figure}

We performed time-resolved experiments at two different Mn thicknesses, corresponding to two different coupling regimes: a strong and a weak parallel coupling regime (Fig.\ \ref{Fig1}(b)). In the weak coupling (WC) regime at a Mn thickness of about 9.5 ML, magnetization reversal proceeds via a two-step process, one relatively sharp at about 15 mT and a broader and smaller one at around 60 mT.  L-MOKE measurements for different Co top-layer thicknesses of 5 ML, 10 ML, and 15 ML reveal a reduction of the coercivity of only the first step with the top Co layer thickness, showing that with increasing field first the top Co (and Ni) layer and then the bottom Co layer reverse magnetization. In the strong coupling (SC) regime at a Mn thickness of about 11 ML, there is a single step in the hysteresis loop because both Co layers reverse their magnetization direction together.  Another WC regime is observed at a Mn thickness of about 8 ML.
Minor loops (Fig.\ \ref{Fig1}(b)) in the WC regime show a horizontal shift of about 3.5 mT in the direction of the magnetic field opposite to the saturation for the reversal of the top Co layer, indicating a parallel interlayer coupling. No antiferromagnetic coupling could be observed in the thickness range of 7--11 ML. In the SC regime, minor loops could not be taken since there is only a single magnetization reversal step. This behavior is consistent with the results from Ref.\ \cite{Zhang.2014}, where the coupling energy between two Co films across a Mn film has been measured for Mn thicknesses from 11 to 17 ML.  Maxima of antiparallel coupling of the direct exchange coupling have been observed at 11.5, 13.5, and 15.5 ML Mn, while an antiparallel coupling maximum of the RKKY coupling was located at 13.7 ML.  Considering the systematic uncertainty of the thickness determination of the Mn wedge of about 10\%, we conclude that at 9.5 ML we are close to an antiparallel maximum of the direct exchange and a parallel maximum of the RKKY coupling.  Together with the magnetostatic interaction favoring parallel alignment, overall the coupling is weakly parallel.  At 11 ML, both the direct exchange and the RKKY coupling favor parallel alignment of the Co layers, leading to the observed strong parallel coupling.  In this case, the magnetizations of both Co layers switch simultaneously if their coercivity is smaller than the coupling field.  

The sample was transferred to the FemtoSlicing Facility at BESSY II \cite{Holldack.2014} within a vacuum suitcase without breaking UHV. There, time-resolved scattering experiments were carried out with a 100\,fs full-width half-maximum (FWHM) circularly polarized soft-X-ray probe pulse and a 60\,fs FWHM 800\,nm $p$-polarized pump pulse from a Ti:Sa laser amplifier system. Both pulses are intrinsically synchronized and propagate close to collinear. The size of the pump laser spot on the sample was (1500 x 200)\,µm² while the size of the X-ray probe pulse was (140 x 40)\,µm². The slicing setup runs with a 6\,kHz repetition rate which is split to record the unpumped and pumped signal in an alternating order for each data point. 
We optimized angle of incidence and photon energy to get the best combination of signal strength and magnetic contrast \cite{Kumberg.2023}. This was the case at the Co $L_3$ edge for an angle of incidence of 7° and an energy of 780.4\,eV.  

An external magnetic field was applied along the X-ray incidence direction, close to in-plane to the sample, to reverse the magnetic order of the FM layers at each data point in order to record the R-XMCD.  Different thicknesses of the Mn wedge were reached by translation of the sample in the vertical direction. R-XMCD hysteresis loops at the Co $L_3$ edge show a one- or two-step reversal in dependence on the Mn thickness, in agreement with our L-MOKE measurements (Fig.\ S3 of the supplemental material \cite{supplement.2025}). 

150\,mT were used for parallel (WCP) alignment of bottom and top Co layers, while a sequence of 150\,mT and a reverse field of 25\,mT was used to set an antiparallel (WCA) alignment in the case of weak coupling at 9.5 ML Mn thickness.  In the case of strong coupling at 11 ML Mn thickness, the magnetization of both Co layers could only be oriented in parallel (SCP).  A pump fluence range of 5--10\,mJ/cm² peak fluence was chosen to vary the range from full demagnetization to considerable demagnetization while still avoiding non-reversible effects like photo-bleaching over the time of the measurements. 

\section{Results}
\label{Results}
We measured the reflected soft-X-ray signal at the Co $L_3$ edge as a function of pump--probe delay time to study the influence of the AFM spacer layer on the ultrafast magnetization dynamics of the coupled Co layers.
For each step of the delay time, pumped and unpumped intensity were recorded for opposite magnetizations of the Co layers $M^+$ and $M^-$, resulting in a total of four signals. The scans were normalized by dividing the difference ($M^+ - M^-$) of the pumped signals by the difference of the unpumped signals, which we define as the normalized difference. Error bars were calculated using error propagation and the total photon count $N$, so that the error scales with $\sqrt{N}$.  Calculations of the layerwise optical differential absorption at 800\,nm wavelength, according to \cite{Ohta.1990, Ohta.1990b}, show an absorption of 16.5$\%$ for the top Co layer and 4.8$\%$ for the bottom Co layer (Fig.\ S6 of the supplemental material \cite{supplement.2025}). Since the top Co layer is 2.5 times as thick as the bottom Co layer, Co atomic moments in both layers should be similarly pumped by the pump pulse and hot electrons will be excited in both layers. 

\begin{figure}[h]
\includegraphics[width=0.6\columnwidth]{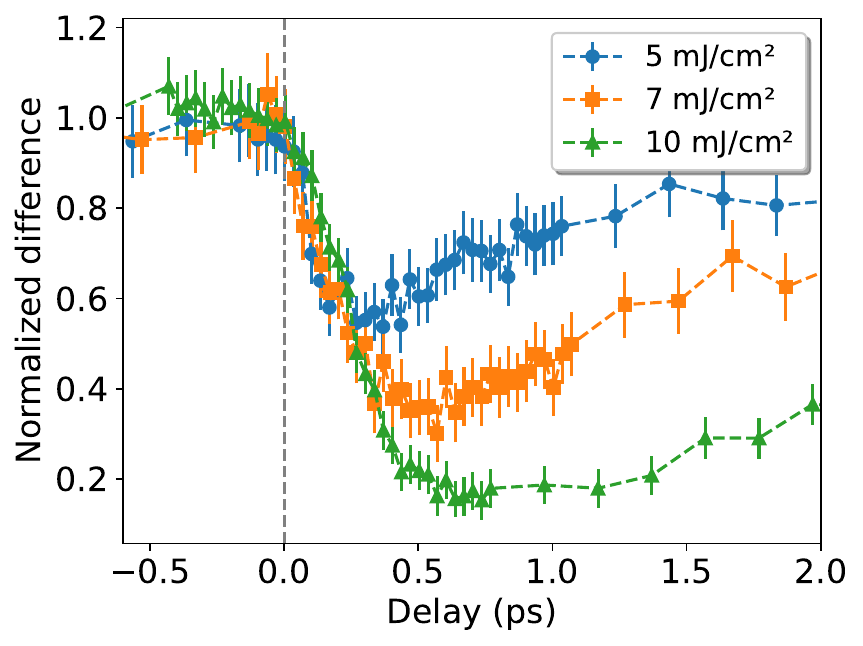}
	\centering
	\caption{Pump--probe delay time traces of the normalized difference for parallel orientation of the Co magnetizations at 9.5 ML Mn thickness at different laser fluences.}	
	\label{Fig4}
\end{figure}

Pump--probe delay scans are shown in Fig.\ \ref{Fig4} for incident pump laser fluences of 5, 7, and 10 mJ/cm$^2$, taken at room temperature with an external field of $\pm$150\,mT to saturate the sample and align both Co layers in parallel.  The normalized difference reaches a minimum within 600\,fs for all three fluences, although a shorter demagnetization time is required for smaller fluences (similar to results from \cite{Kumberg.2020} and \cite{Koopmans.2010}). 
The demagnetization amplitude increases with increasing fluence, resulting in a nearly complete demagnetization at a fluence of 10 mJ/cm². Remagnetization, indicated by an increase of normalized difference with delay time, takes longer for larger pump fluences as more energy needs to be dissipated to return to equilibrium, and is in agreement with previous observations \cite{Roth.2012}.

To evaluate the magnetization dynamics quantitatively, we fitted our measurements to the sum of three exponential functions.  One exponential describes the ultrafast demagnetization and the other two a fast initial and a slower subsequent remagnetization.  Additionally, the sum is convoluted by a Gaussian $g(t)$ with a FWHM of 120\,fs to consider the temporal laser profile. This results in the following equation:
\begin{dmath}
\label{triple fit}
\dfrac{M}{M_0}(t) = g(t) \otimes \left[\Theta(t-t_0) \left(a(\mathrm{e}^{-(t-t_0)/t_m}-1)\\-b(\mathrm{e}^{-(t-t_0)/t_f}-1)-c(\mathrm{e}^{-(t-t_0)/t_s}-1)\right)+C\right]
\end{dmath}
where $\Theta$ is the Heaviside step function, $t_m$, $t_f$, and $t_s$ the time constants for demagnetization and fast and slow remagnetization, and $a$, $b$, and $c$ are the corresponding de- and remagnetization amplitudes, respectively. \textit{C} is an offset to start the fitting function before time zero at the same value as the normalized measurement data. Time $t_0$ is a shift to accommodate for slow drifts in pump--probe delay time between measurements to have a common time zero for all fluences. (Overall, for long time windows ($\sim$ 100\,ps), remagnetization amplitudes $b$ and $c$ should be equal to the demagnetization amplitude $a$, which was not enforced but usually the case during fitting.)

We study the influence of the AFM spacer layer by recording pump--probe delay scans at the two different selected Mn thicknesses to compare the dynamics when both Co layers are weakly and strongly coupled (SCP). Additionally, in the case of weak coupling at 9.5 ML of Mn, we looked at ultrafast magnetization dynamics in the case of antiparallel (WCA) and parallel (WCP) alignment of top and bottom Co layers. In the WCP case, Mn spins need to twist within
the AFM spacer layer to accommodate both Co layers, since the direct exchange coupling between Mn and Co favors an antiparallel alignment.

\begin{figure}[h]
\includegraphics[width=0.6\columnwidth]{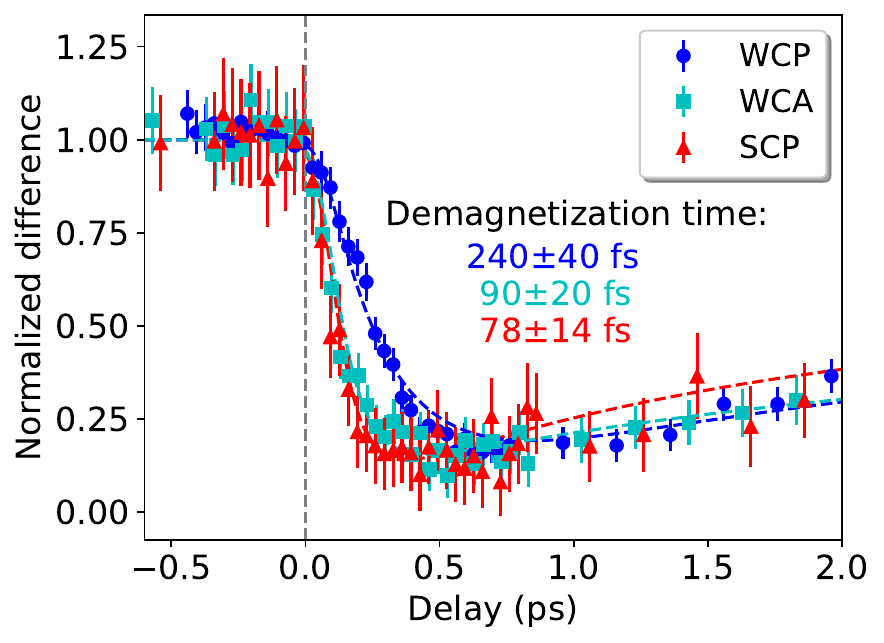}
	\centering
	\caption{Pump--probe delay time traces for a fluence of 10\,mJ/cm² for WCP (blue), WCA (cyano) and SCP (red). The dashed lines are the corresponding fit functions.}	
	\label{APvsP}
\end{figure}

 Delay scans and fit results are presented in Fig. \ref{APvsP} for all three cases for a pump fluence of 10\, mJ/cm². Extracted time constants for pump fluences of 5, 7, and 10 mJ/cm² are listed in Tab.\,\ref{tab:triple fit results}. Comparing WCP and WCA (Fig.\ \ref{APvsP} blue and cyano), we measure a significantly faster demagnetization of $90\pm20\,\mathrm{fs}$ in the antiparallel case compared to $240\pm40\,\mathrm{fs}$ in the parallel case for a fluence of 10\,mJ/cm². 

In the case of SCP (Fig.\ \ref{APvsP} red) at 11\,ML of Mn, where only parallel alignment of both Co layers is achievable, a fast demagnetization of $78\pm14\,\mathrm{fs}$ was measured, with a value in the range of the WCA case. This is quite surprising, as one would assume a slower demagnetization for parallel alignment in the case of superdiffusive spin currents, which thus cannot explain this behavior and will be discussed in the following. 

\begin{table}[htbp]
  \centering
    \begin{tabular}{c|c|c|c|c}
      Fluence    &   Alignment   & $t_m$ [ps] & $t_f$ [ps] & $t_s$ [ps] \\
      & WCP & 0.21 $\pm$ 0.06 & 0.8 $\pm$ 0.28 & 16 $\pm$ 10 \\
    5  mJ/cm² & WCA & 0.11 $\pm$ 0.04 & 0.5 $\pm$ 0.3 & 5 $\pm$ 3 \\
          & SCP & 0.11 $\pm$ 0.04 & 0.5 $\pm$ 0.2 & 6 $\pm$ 4 \\
          &       &       &       &  \\
      & WCP & 0.17 $\pm$ 0.05 & 0.8 $\pm$ 0.3 & 8 $\pm$ 2 \\
    7 mJ/cm² & WCA & 0.10 $\pm$ 0.03 & 0.71 $\pm$ 0.23 & 13 $\pm$ 3 \\
          & SCP & 0.08 $\pm$ 0.02 & 1.1 $\pm$ 0.3 & 17 $\pm$ 7 \\
          &       &       &       &  \\
      & WCP & 0.24 $\pm$ 0.04 & 2.0 $\pm$ 0.5 & 110 $\pm$ 40 \\
    10 mJ/cm² & WCA & 0.09 $\pm$ 0.02 & 2.1 $\pm$ 0.5 & 210 $\pm$ 160 \\
          & SCP & 0.08 $\pm$  0.01 & 1.6 $\pm$ 0.3 & 60 $\pm$ 20 \\
    \end{tabular}
    \caption{Demagnetization time constant $t_m$ as well as fast $t_f$ and slow $t_s$ remagnetization time constants in dependence of pump laser fluence in case of parallel (WCP, SCP) or antiparallel (WCA) alignment of both Co layers. For SC, only parallel alignment (SCP) is possible due to strong interlayer coupling.} 
  \label{tab:triple fit results}
\end{table}

\section{Discussion}
\label{Discussion}
\begin{figure}[H]
\includegraphics[width=0.75\columnwidth]{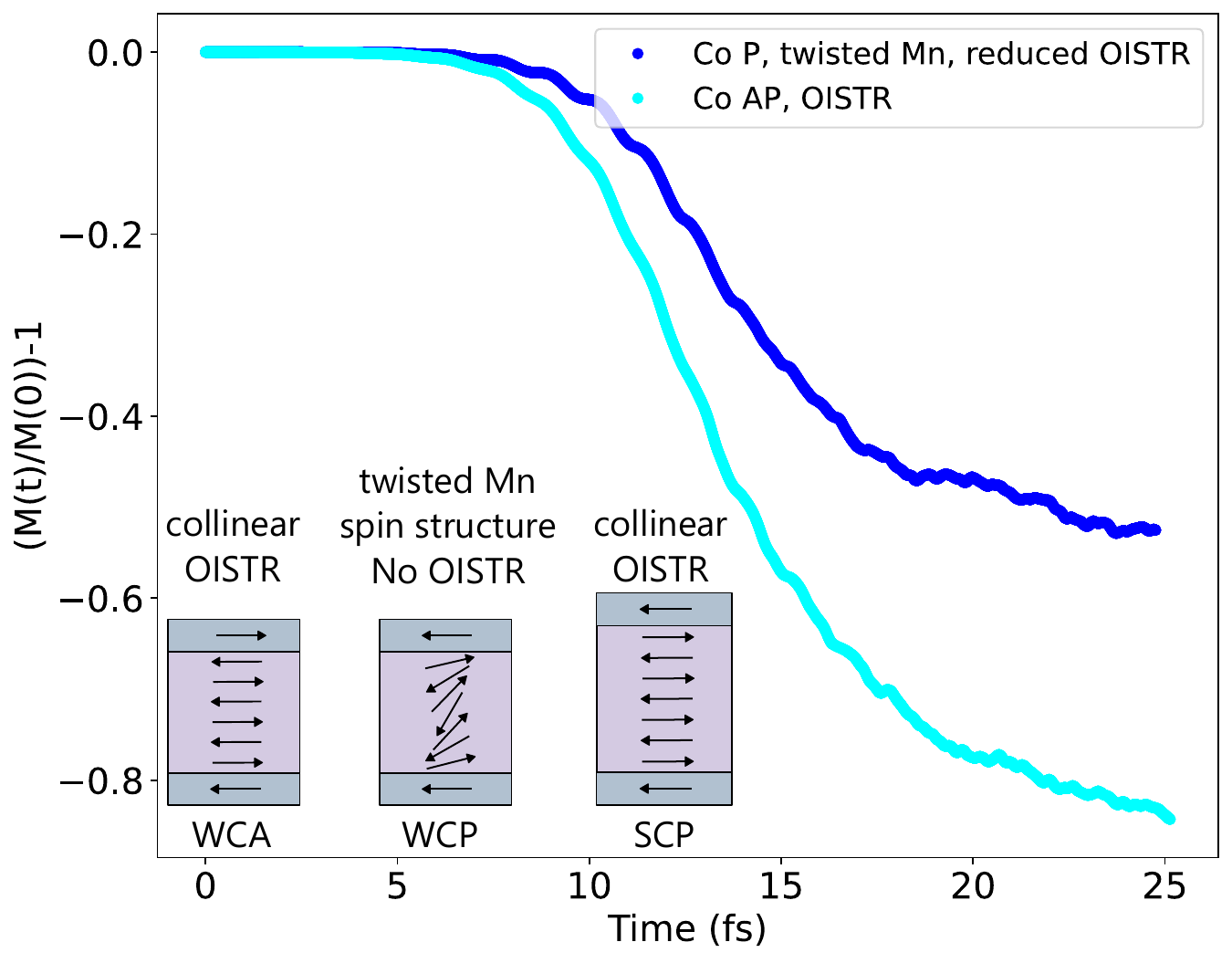}
	\centering
	\caption{Result of a TD-DFT calculation of the time-dependent magnetic moment of Co for a Co--Co--Mn--Mn--Mn--Mn--Co--Co layer system after excitation with a 12.4\,fs FWHM 800\,nm laser pulse with a fluence of 29\,mJ/cm² in the case of antiparallel (cyano) and parallel (blue) alignment of Co spins on opposite sides of the trilayer. The graph shows the change in the sum of the absolute value of the magnetic moments of the Co layers on both sides of the Mn spacer layer.  The differences in demagnetization dynamics and amplitude of the Co layers for parallel and antiparallel alignment appear due to OISTR with the sandwiched Mn layer. Inset: Schematic of Mn spin structure for the WCP, WCA and SCP case. Direct coupling of Mn spins with Co spins causes Mn spins within the layer to twist in the case of WCP, such that no or reduced OISTR can happen between Co and Mn atoms.}	
	\label{Fig6}
\end{figure}

Since the accelerated ultrafast demagnetization of the Co layers takes place within the first 100--200 fs after pumping the system, faster than the typical timescale for equilibration with the lattice, the effect seems to be optical in nature.  
As mentioned before, the exchange of spin currents between the two FM layers would explain the different demagnetization times for parallel and antiparallel alignment of the FM layers in the WCP and WCA case, but cannot explain the observed accelerated demagnetization in the SCP case.  We thus attribute the difference in demagnetization times to the spin structure of the AFM Mn spacer layer:  In the case of WCA and WCP, the direct exchange coupling across the Mn layer favors antiparallel alignment of the adjacent Co layers.  Thus, there is a low-energy collinear Mn spin structure in the case of antiparallel alignment of both Co layers, while for parallel alignment, Mn spins need to twist within the layer to accommodate both Co layers to which they couple directly at the interface, increasing the exchange energy.  In the case of SCP, the direct exchange favors parallel alignment, such that the AFM spin structure rests in a collinear configuration.  This is schematically depicted in the inset of Fig.\ \ref{Fig6}.  Figure \ref{APvsP} and Tab.\ \ref{tab:triple fit results} thus show a clear dependence of the time constant of the ultrafast optical demagnetization of the Co layers on the Mn spin structure.  Because of the ultrafast nature of the effect, we attribute this observation to the OISTR effect. 

To support this assumption and to explore this phenomenon qualitatively, we use TD-DFT, based on a fully non-collinear version of the Elk code \cite{Dewhurst.2016, http:elk.sourceforge.net.}. We model the system by a stack of 2 ML Co, 4 ML Mn, and 2 ML Co. Since OISTR is an effect that is most prominent between nearest neighbors, extra layers are not needed to unravel the necessary physics \cite{Dewhurst.2018}. We calculate the dynamic optical response after excitation with a pump pulse of 29\,mJ/cm² fluence with a wavelength of 800\,nm and a FWHM of 12.4\,fs, which is shorter than the employed 60\,fs pump pulse, but should show the same effects, as shown in \cite{Siegrist.2019}. Figure \ref{Fig6} shows the time-dependent relative change of the sum of the absolute values of the magnetic moments of both Co layers after excitation with a pump pulse for alignment of the Co magnetizations in parallel (P, blue) and antiparallel (AP, cyano). The calculations clearly show that there is indeed a large difference in dynamics between parallel and antiparallel alignment of both Co layers coming from OISTR between Co and Mn atoms.  For the idealized system considered in the calculations, the even number of Mn atomic monolayers means a collinear spin structure for antiparallel Co magnetizations and a twisted one for parallel. The OISTR mechanism is strongly governed by the number of available states, which is large (maximum possible) for a fully AFM configuration between Co and Mn. In a twisted non-collinear spin configuration this number of available states, projected in the direction of Co spins, automatically reduces, reducing the OISTR effect. 

This can be transferred to explain our experimental results, as shown in the inset of Fig.\ \ref{Fig6}: the presence of OISTR in the WCA case and in the SCP case result in an accelerated ultrafast demagnetization of the Co layers.  For WCP, the direct exchange coupling through the Mn layer leads to a twisted Mn spin structure. As a result, there is no or less OISTR in the latter configuration.  
The mechanism in OISTR is that unoccupied minority states in Mn can act as a spin sink for majority spins from Co. The demagnetization time decreases due to optical transfer of spin-polarized electrons from both Co layers into Mn, which acts as a decay channel. Both, the calculations and the experiments confirm that the spin structure of the AFM spacer layer influences significantly the magnetization dynamics of adjacent FM layers after an ultrafast optical excitation.  
We note that spin-dependent interface transmittance of hot electrons from the Co layers into the sandwiched Mn layer, assuming a reduced spin penetration depth in the case of twisted Mn spin structure, could be another explanation.  In \cite{Malinowski.2008} and \cite{Rudolf.2012}, an accelerated demagnetization in the antiparallel case has been observed and explained via superdiffusive spin currents. However, in both cases only a relatively low pump fluence (0.65\,mJ/cm² and 2\,mJ/cm², respectively) was used to avoid competing processes, because at high pump fluences superdiffusive spin currents become saturated or even less effective close to full demagnetization \cite{Battiato.2010, Rudolf.2012, Battiato.2012}. In our case, larger pump fluences were used, at which the OISTR effect becomes relevant and the effect of superdiffusive spin currents negligble. 
Differences in interface quality between the top Co layer and Mn could in principle also change the magnetization dynamics between thin (9.5\,ML) and thick (11\,ML) Mn spacer layers, but firstly we expect such differences to be relatively small, and secondly they do not play a role when changing between parallel and antiparallel alignment of the Co magnetization directions at fixed Mn thickness.
The capping Ni layer could have an influence on the dynamics as well, but it should influence all three cases equally and we expect it to behave identically to the top Co layer, as the R-XMCD hysteresis loop (Fig.\ S5 of the supplemental material \cite{supplement.2025}) indicates.

\section{Conclusion}
\label{Conclusion}
In summary, we probed element- and time-resolved ultrafast magnetization dynamics of two FM layers, magnetically coupled through an AFM spacer layer. From our measurements it is clear that the antiferromagnetic spacer layer and its spin structure have a significant effect on the magnetization dynamics of the FM layers after an optical pump. We show that the experimental results can be qualitatively reproduced by TD-DFT calculations and attribute their origin to the OISTR effect, where spin-selective optical excitation of electrons from occupied Co states into unoccupied Mn states act as an additional decay channel to enhance and accelerate demagnetization. Our findings provide further insight into the fundamental mechanisms that play a role in ultrafast magnetization dynamics. Accelerating demagnetization may be used in applications to improve dynamics for fast optical switching of magnetic order for spintronic devices or data storage.

\vspace{-5mm}
\begin{acknowledgments}
	This work was supported by the Deutsche Forschungsgemeinschaft via the CRC/TRR 227 ``Ultrafast Spin Dynamics", project-ID: 328545488, projects A03, A04, and A07. We thank the Helmholtz-Zentrum Berlin f\"ur Materialien und Energie for the allocation of synchrotron radiation beamtime, and M. A. Mawass for assistance in using the FemtoSlicing Facility.
\end{acknowledgments}

%\bibliography{Literaturverzeichnis}


\end{document}

\title{Accelerated ultrafast demagnetization of an interlayer-exchange-coupled Co/Mn/Co trilayer 
\\--- Supplemental Material ---}

\author{Jendrik G\"ordes}
\affiliation{Institut f\"ur Experimentalphysik, Freie Universit\"at Berlin, Arnimallee 14, 14195 Berlin, Germany}
\author{Ivar Kumberg}
\affiliation{Institut f\"ur Experimentalphysik, Freie Universit\"at Berlin, Arnimallee 14, 14195 Berlin, Germany}
\author{Chowdhury S. Awsaf}
\affiliation{Institut f\"ur Experimentalphysik, Freie Universit\"at Berlin, Arnimallee 14, 14195 Berlin, Germany}
\author{Marcel Walter}
\affiliation{Institut f\"ur Experimentalphysik, Freie Universit\"at Berlin, Arnimallee 14, 14195 Berlin, Germany}
\author{Tauqir Shinwari}
\affiliation{Institut f\"ur Experimentalphysik, Freie Universit\"at Berlin, Arnimallee 14, 14195 Berlin, Germany}
\author{Sangeeta Thakur}
\affiliation{Institut f\"ur Experimentalphysik, Freie Universit\"at Berlin, Arnimallee 14, 14195 Berlin, Germany}
\author{Sangeeta Sharma}
\affiliation{Institut f\"ur Experimentalphysik, Freie Universit\"at Berlin, Arnimallee 14, 14195 Berlin, Germany}
\affiliation{Max-Born-Institut f\"ur Nichtlineare Optik und Kurzzeitspektroskopie, Max-Born Stra{\ss}e 2a, 12489 Berlin}
\author{Christian Sch\"u{\ss}ler-Langeheine}
\affiliation{Helmholtz-Zentrum Berlin f\"ur Materialien und Energie, Albert-Einstein-Stra{\ss}e 15, 12489 Berlin}
\author{Niko Pontius}
\affiliation{Helmholtz-Zentrum Berlin f\"ur Materialien und Energie, Albert-Einstein-Stra{\ss}e 15, 12489 Berlin}
\author{Wolfgang Kuch}
\affiliation{Institut f\"ur Experimentalphysik, Freie Universit\"at Berlin, Arnimallee 14, 14195 Berlin, Germany}

\date{\today}

\renewcommand{\thefigure}{S\arabic{figure}}
\renewcommand{\theequation}{S\arabic{equation}}
\renewcommand{\thetable}{S \Roman{table}}

\maketitle

\subsection*{Sample growth}
\begin{figure}[h]
\includegraphics[scale=0.5]{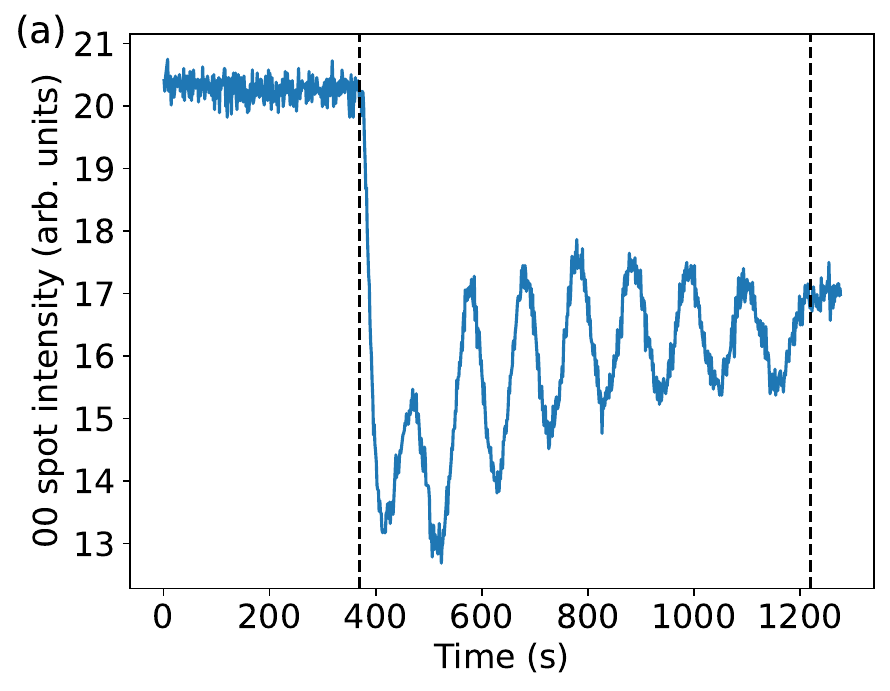}
\includegraphics[scale=0.5]{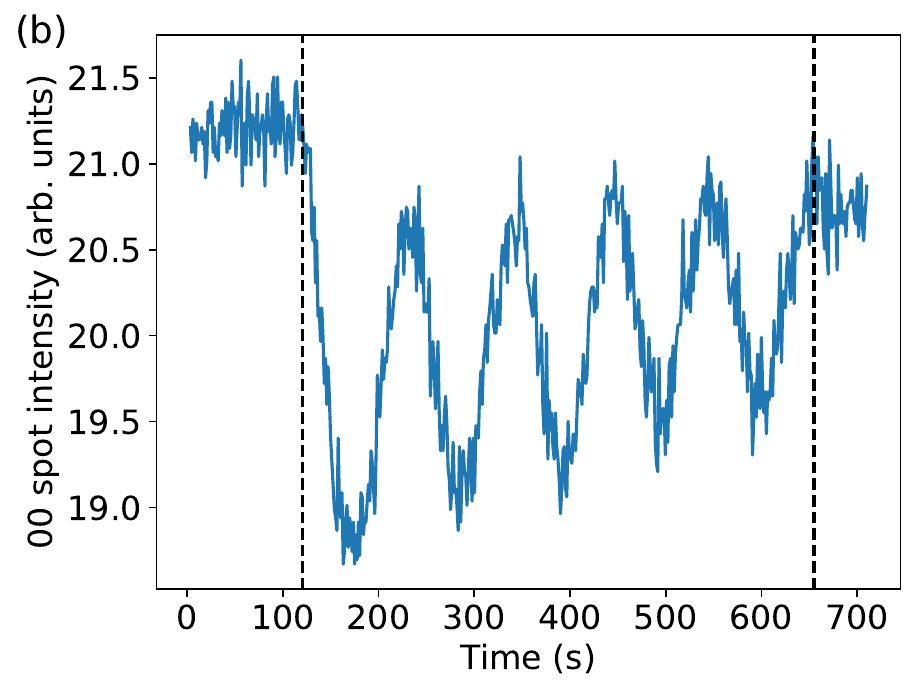}
	\caption{MEED intensity oscillation of the (00) spot during evaporation of 8\,ML Co on Cu(001) (a) and 5\,ML Co on (9$\pm$1)\,ML Mn/8\,ML Co/Cu(001) (b). Beginning and end of evaporation are indicated by vertical dashed black lines.}	
	\label{FigS1}
\end{figure}

\begin{figure}[h]
\includegraphics[scale=0.5]{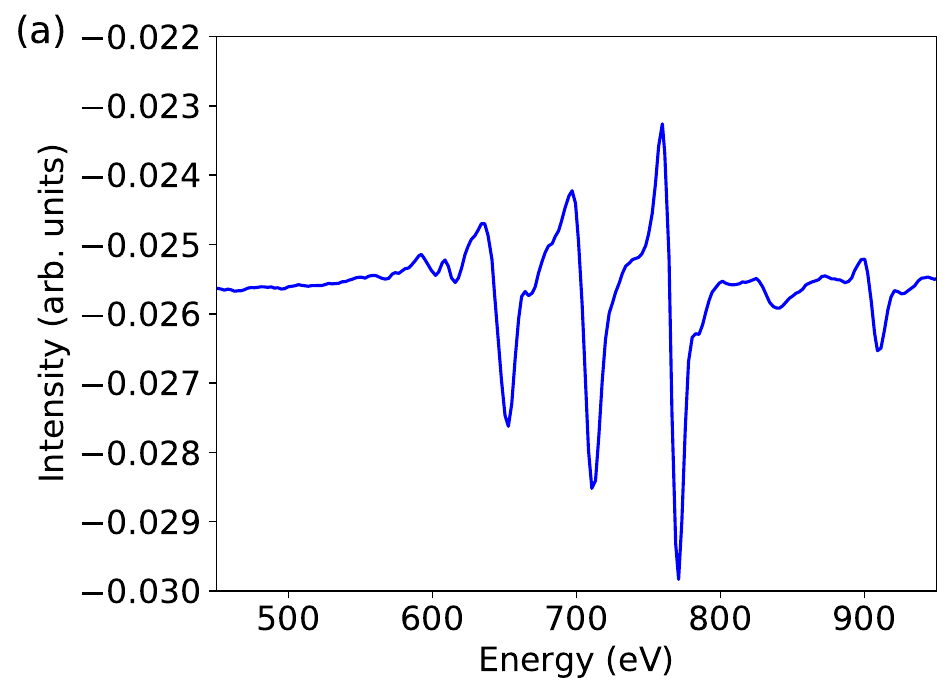}
\includegraphics[scale=0.5]{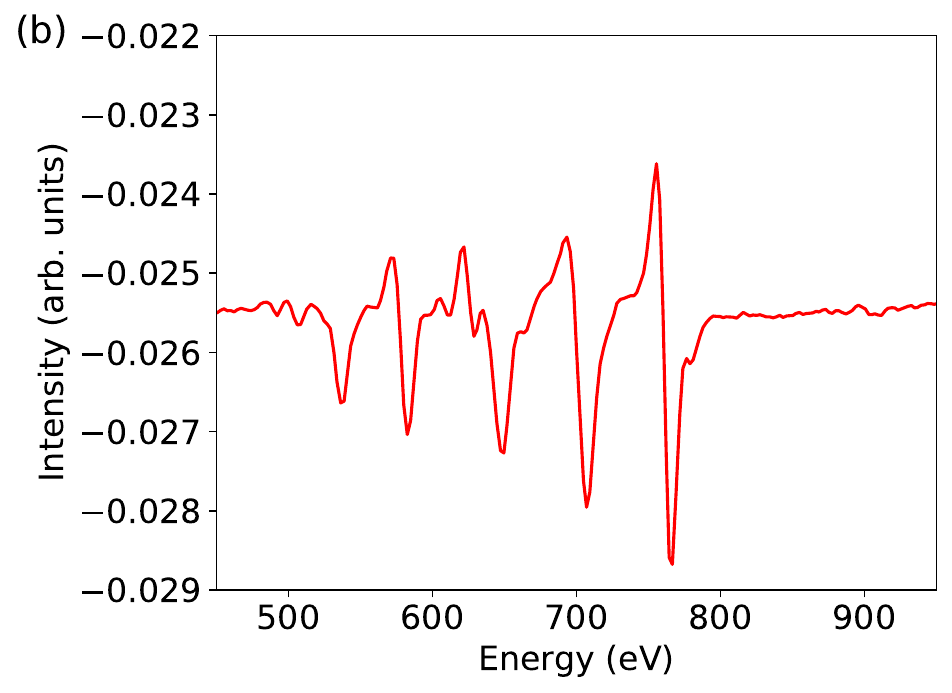}
	\caption{AES spectra of 8\,ML of Co on Cu(001) (a) and 5\,ML Co on (9$\pm$1)\,ML Mn/8\,ML Co/Cu(001) (b).}
	\label{FigS2}
\end{figure}

Layer thicknesses were monitored during evaporation by a quartz crystal microbalance (QCM) as well as by intensity oscillations of the (00) spot of medium energy electron diffraction (MEED). MEED oscillations for the first 8\,ML of the bottom Co layer on Cu(001) substrate as well as the first 5\,ML of the top Co layer on Mn are shown in Fig.\,\ref{FigS1}.
\\
The corresponding spectra from Auger electron spectroscopy (AES) are displayed in Fig.\,\ref{FigS2} to check the film thickness. AES was also taken after 10\,ML, 15\,ML and 20\,ML of top Co layer evaporation and after Ni evaporation, but not after Mn evaporation to avoid oxidation of the Mn layer. In all AES spectra there was no significant oxidation or other contamination detected.

\subsection*{R-XMCD hysteresis loops at Co \textit{L}\textsubscript{3} and Ni \textit{L}\textsubscript{3} edge}
\begin{figure}[h]
\includegraphics[scale=0.5]{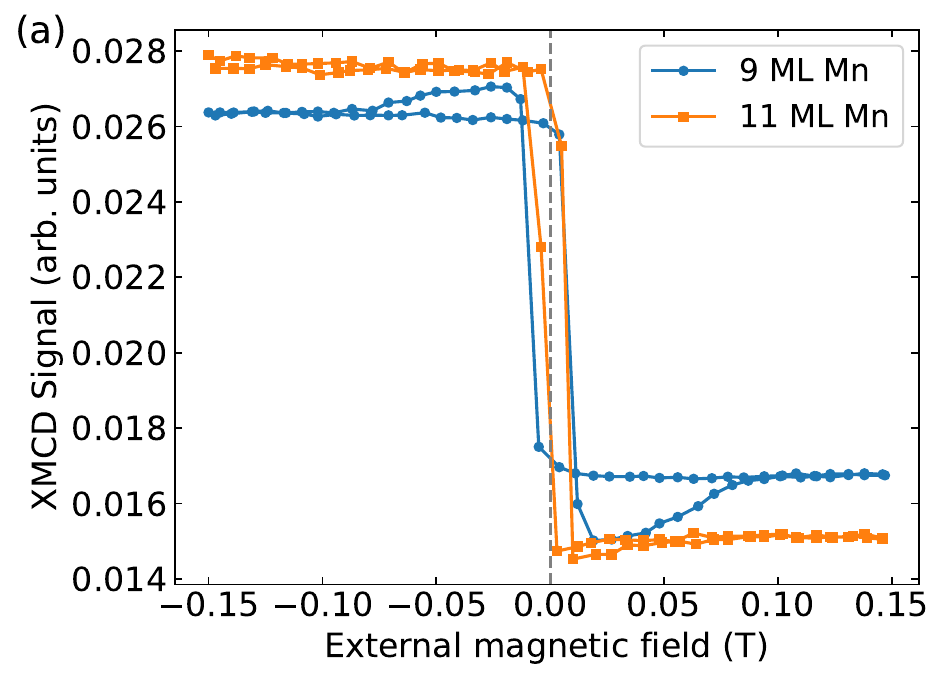}
\includegraphics[scale=0.5]{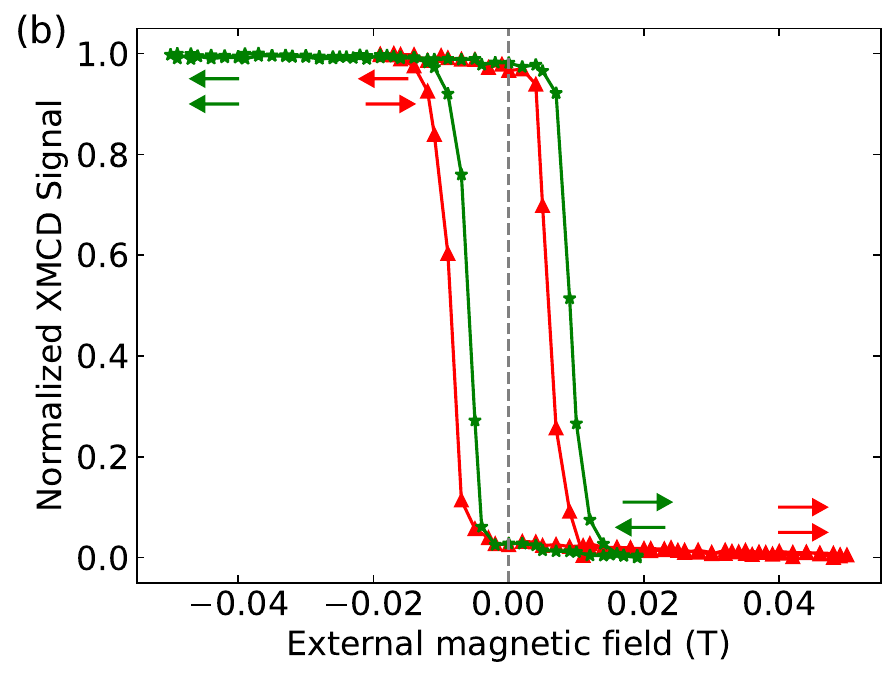}
	\caption{(a): R-XMCD hysteresis loops at the Co $L_3$ edge for weak (9 ML Mn, blue) and strong (11 ML Mn, orange) coupling. (b): Minor loops for a Mn spacer layer thickness of 9\,ML. A positive horizontal shift indicates a preferential parallel interlayer coupling of both Co layers.}	
	\label{FigS3}
\end{figure}

\begin{figure}[h]
\includegraphics[scale=0.5]{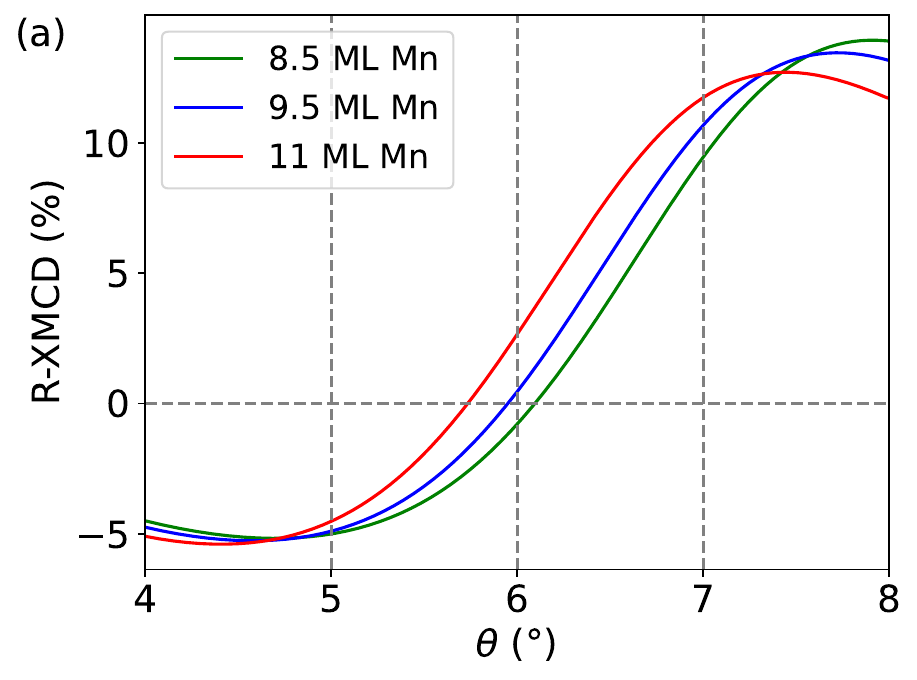}
\includegraphics[scale=0.5]{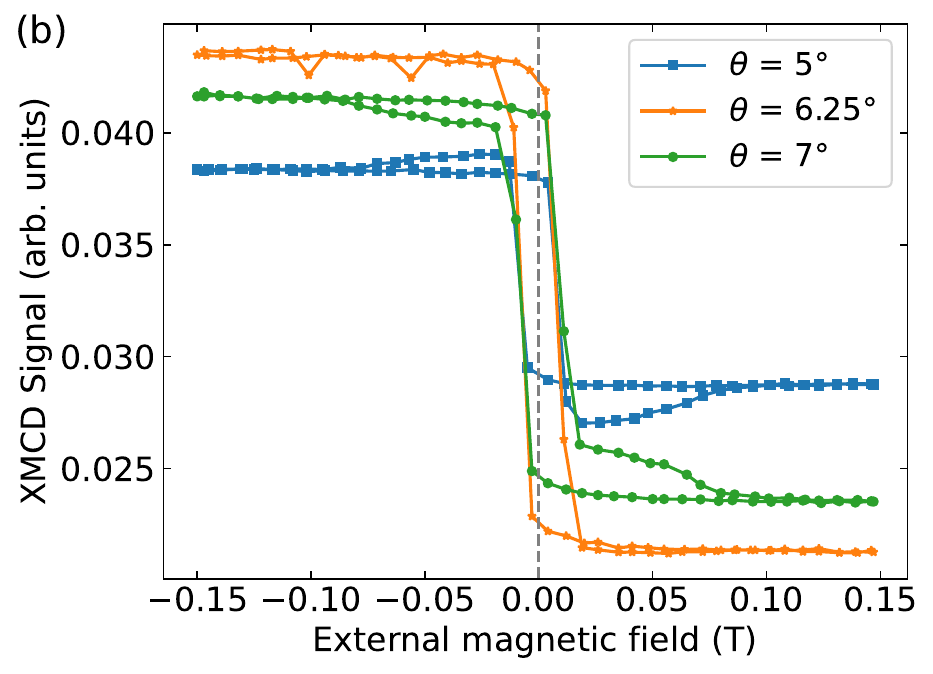}
	\centering
	\caption{(a): \textit{ReMagX} simulations of the R-XMCD of both Co layers for different Mn spacer layer thicknesses. Increasing the incidence angle $\theta$ from 5° to 7° changes the contribution of the bottom Co layer to the R-XMCD signal from negative to positive around a value of about 6°. (b): R-XMCD hysteresis loops, taken at the Co $L_3$ edge, for 9 ML Mn thickness and different incidence angles $\theta$. The relative contribution of the bottom Co layer $R$ to the R-XMCD signal changes sign between 5° and 7° and is negligible for 6.25°.}	
	\label{FigS4}
\end{figure}

\begin{figure}[h]
\includegraphics[scale=0.5]{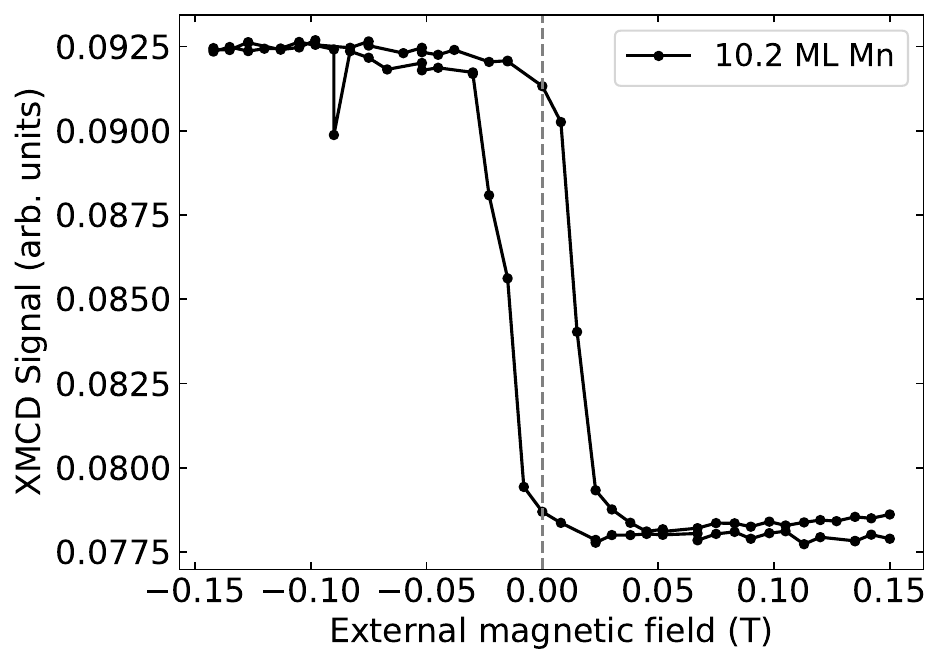}
	\centering
	\caption{Hysteresis loop of the sample, measured at the Ni $L_3$ edge with an incidence angle $\theta$ of 6.3° for a Mn spacer layer thickness of 10.2 ML.}
	\label{FigS5}
\end{figure}

Reflectivity at the Co $L_3$ edge, measured at the UE56-1-ZPM beamline at BESSY II, shows a clear R-XMCD upon reversal of the external magnetic field of 150 mT. 
R-XMCD hysteresis loops at the Co $L_3$ edge for different Mn thicknesses (Fig.\ \ref{FigS3} (a)) show a one- or two-step reversal in agreement with our L-MOKE measurements (see Fig.\ 1 (b) in main text). Minor loops reproduce the positive horizontal shift in the weak coupling regime (Fig.\ \ref{FigS3} (b)).
The negative contribution to the R-XMCD of the bottom Co layer in Fig.\ \ref{FigS4} (a) can be rationalized as a consequence of the contribution of the real part of the optical constant \cite{Kumberg.2023}.  Changing the incidence angle of the incoming X-ray beam from 7° to 5° results in a change of the hysteresis loop shape (Fig. \ref{FigS4} (b)). The contribution of the bottom Co layer reverses sign between 7° and 5° and crosses zero at 6.25°. While the second step of the hysteresis loop reduces the R-XMCD signal at an angle of 5°, it enhances the signal at 7° and disappears at 6.25°.  Simulations of the angle-dependent X-ray reflectivity with the software package \textit{ReMagX} \cite{Macke.2014, https:www.remagx.org.} for our layer system support this assumption (for further information see below). Figure \ref{FigS4} (a) shows the difference between simulated R-XMCD signals at the Co $L_3$ edge for opposite bottom-layer magnetization as a function of $\theta$. There is a change of sign of the contribution of the bottom layer at an angle of around 6°, depending weakly on Mn thickness. Consequently, the hysteresis loop taken at $\theta = 6.25$° shows only the contribution of the top layer, resulting in a single-step magnetization reversal process. L-MOKE measurements do not show this behavior, as the contribution of the bottom Co layer is always positive. 
\\
Figure \ref{FigS5} shows the hysteresis loop of the sample at the Ni $L_3$ edge at a photon energy of 854\,eV and a Mn thickness of 10.2 ML. The magnetic field needed to reach saturation of the Ni layer is around 25\,mT, similar to the one required for the top Co layer.

\subsection*{ReMagX simulation}
\textit{ReMagX} is a simulation program for X-ray reflectivity measurements which takes into account magnetic contributions from the layers of the system. Here, we used the Paratt algorithm with circularly polarized light. To consider the contribution of the bottom Co layer on the magnetic dichroism, its sign of magnetization was changed and the difference in XMCD reflectivity (between antiparallel and parallel configuration) was calculated for different incidence angles $\theta$. Delta, beta, delta-m and beta-m values were calculated for Co, Ni and Mn from XAS and XMCD data taken at the PM3 beamline of BESSY II from a $\mathrm{Cu_3Au(100)/[1\,ML\,Ni/3\,ML\,Mn]_4/10\,ML\,Co}$ sample. Cu values were taken from Henke tables \cite{Henke.1993}, since it is at off-resonant energy and has no magnetic contribution. The X-ray energy was set to 780.5\,eV, identical to the energy used during the measurements. As vertical layer distances, 1.76\,\r{A}/ML for Ni \cite{Abujoudeh.1986}, 1.74\,\r{A}/ML for Co \cite{Cerda.1993} and 1.89\,\r{A}/ML for Mn \cite{Wang.1995} were used. 

\subsection*{Absorption profile}
\begin{figure}[H]
\includegraphics[scale=0.5]{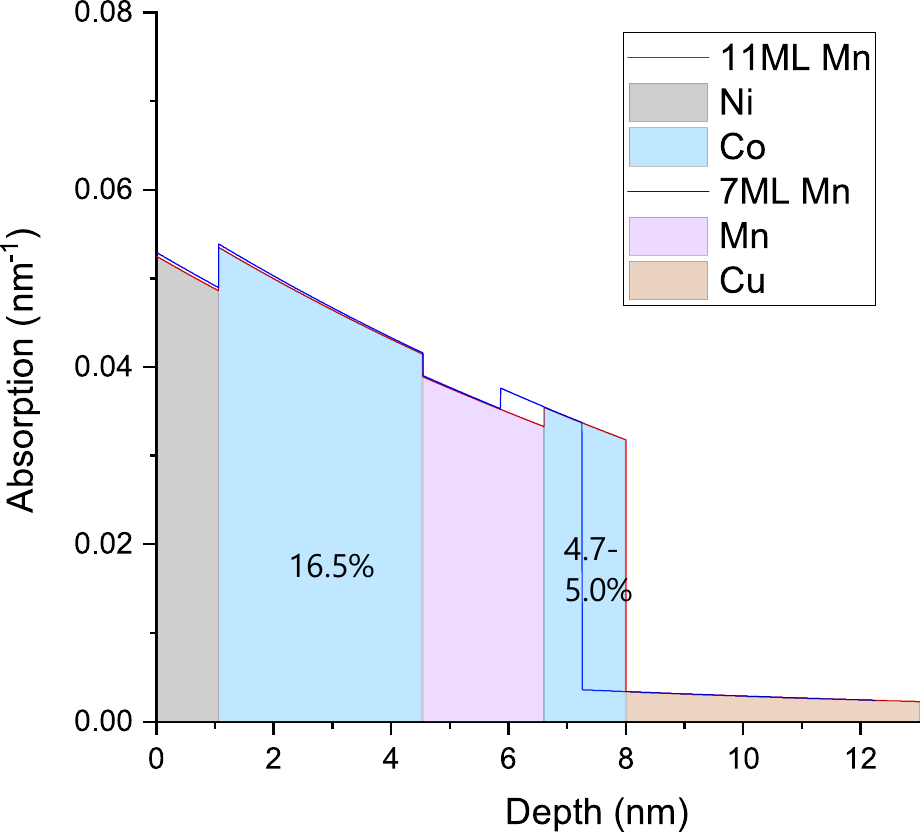}
	\centering
	\caption{Layer-dependent differential absorption profile for an 800\,nm pump pulse in the case of a Cu(001)/8\,ML Co/$x$-Mn/20\,ML Co/6\,ML Ni sample. Two profiles are plotted for a Mn thickness of 7\,ML (blue line) and 11\,ML (red line) resulting in an absorption of 16.5\,\% for the top Co layer and 4.7--5.0\,\% for the bottom Co layer of the pump-pulse fluence. For the semi-infinite Cu(001) substrate, only the first 5\,nm are shown.}	
	\label{FigS6}
\end{figure}

The layerwise absorption profile in Fig.\ \ref{FigS6} was calculated by using the matrix formalism expressed in \cite{Ohta.1990, Ohta.1990b}. Just as for the \textit{ReMagX} simulations, vertical interlayer distances with a value of 1.76\,\r{A}/ML for Ni, 1.74\,\r{A}/ML for Co and 1.89\,\r{A}/ML for Mn were used. Here, we calculated the absorption both for a Mn thickness of 7\,ML and 11\,ML. For a pump pulse with 800\,nm wavelength, optical constants of Ni, Co, Mn and Cu were taken from \cite{Johnson.1974} and are shown in Table\,\ref{tab:S1}.
\begin{table}[htbp]
  \centering
  \caption{Parameters used for the calculation of the layerwise absorption. The absorption (abs.) in \% is the total absorption of the incoming light in the whole layer.}
    \begin{tabular}{|c|c|c|c|c|}
    \hline
    Layer & Refractive index & thickness\,[nm] & abs.\,[\%]\,(7 ML Mn) & abs.\,[\%]\,(11 ML Mn) \\
    \hline
    Vacuum &1&-&-&-\\
    Ni & 2.22 + 4.89$i$ & 1.1 & 5.4 & 5.4 \\
    Co & 2.49 + 4.80$i$ & 3.5 & 16.5 & 16.4 \\
    Mn & 2.79 + 3.99$i$ & 1.3 or 2.1 & 4.9 & 7.5 \\
    Co & 2.49 + 4.80$i$ & 1.4 & 5.0 & 4.7 \\
    Cu & 0.2535 + 5.013$i$ & 1$\cdot$10e6 & 4.5 & 4.2 \\
    Vacuum & 1 & - & - & - \\
    \hline
    \end{tabular}
  \label{tab:S1}
\end{table}

%\bibliography{Literaturverzeichnis}
